%% file: main.tex
\documentclass[ba]{imsart} 

\usepackage{etoolbox}
\newbool{arxiv}

\boolfalse{arxiv} 

\usepackage{appendix}

\usepackage{amsthm}
\usepackage{amsmath}
\usepackage{natbib}
\usepackage[colorlinks,citecolor=blue,urlcolor=blue,filecolor=blue,backref=page]{hyperref}
\usepackage{graphicx}

\usepackage{microtype}
\usepackage{graphicx}
\usepackage{subfigure}
\usepackage{booktabs} 
\usepackage{siunitx}
\usepackage{hyperref}
\usepackage{xargs}[2008/03/08]
\usepackage{xfrac}

\usepackage{blindtext}
\usepackage[inline]{enumitem}

\usepackage{cleveref}

\usepackage{amssymb}
\usepackage{amsfonts}
\usepackage{mathtools}
\usepackage{colonequals}
\usepackage{algpseudocode, algorithm} 
\usepackage{xargs} 

\usepackage{mathrsfs}  

\usepackage{pdfpages} 

\notbool{arxiv}{
    \startlocaldefs
}

\input{_math_defs.tex}

\def\maceachern{\textbf{MacEachern and Lee}}
\def\griffin{\textbf{Griffin and Kalli}}
\def\gilleyva{\textbf{Gil-Leyva and Mena}}
\def\ascolani{\textbf{Ascolani, Catalano, and Pr{\"u}nster}}
\def\rebaudo{\textbf{Rebaudo, Fasano, Franzolini, and M{\"u}ller}}
\def\miscouridou{\textbf{Miscouridou and Panero}}

\notbool{arxiv}{
    \endlocaldefs
}

\notbool{arxiv}{
    \pubyear{2023}
    \volume{TBA}
    \issue{TBA}
    \firstpage{1}
    \lastpage{13}
}

\begin{document}

\title{Evaluating Sensitivity to the Stick-Breaking Prior in Bayesian Nonparametrics (Rejoinder)}

\ifbool{arxiv}{
    \author{
      \hspace{4em}
      Ryan Giordano \thanks{Equal contribution author}
      \thanks{Department of EECS, MIT}

      \and
      Runjing Liu \footnotemark[1]
      \thanks{Department of Statistics, UC Berkeley}

      \hspace{4em}
      \and
      Michael I.~Jordan \footnotemark[3]
      \and
      Tamara Broderick \footnotemark[2]
    }

    \maketitle

    \input{abstract}
} {
    \runtitle{Evaluating Sensitivity to the Stick-Breaking Prior in BNP}
    \runauthor{Giordano, Liu, Jordan, and Broderick}

    \newcommand\authors{
        \begin{aug}
            \author{\fnms{Ryan} \snm{Giordano}\thanksref{addr1,t1}},
            \author{\fnms{Runjing} \snm{Liu}\thanksref{addr2,t1}},
            \author{\fnms{Michael I.} \snm{Jordan}\thanksref{addr2}},
            \and
            \author{\fnms{Tamara} \snm{Broderick}\thanksref{addr1}}

            \address[addr1]{Department of EECS, MIT,
            77 Massachusetts Ave., 38-401,
            Cambridge, MA 02139}

            \address[addr2]{Department of Statistics,
            367 Evans Hall, UC Berkeley,
            Berkeley, CA 94720}

            \thankstext{t1}{Equal contribution. }
        \end{aug}
    }

    \begin{frontmatter}
    \authors{}

    \end{frontmatter}
}

%
%
%
%
%
%

\section{Introduction}
\input{introduction}

\section{What does it take to do local robustness?}\label{sec:derivatives}
\input{derivatives}

\subsection{Some requested derivatives}\label{sec:requested}
\input{requests}

\section{What makes a derivative useful?}\label{sec:useful}
\input{useful}

\typeout{}

\ifbool{arxiv} {
    \bibliographystyle{plainnat}
} {
    \bibliographystyle{ba}
}
\bibliography{references}




\end{document}

%% file: _math_defs.tex
\def\ind#1{\mathbb{I}\left(#1\right)}

\def\fracat#1#2#3{\left.\frac{#1}{#2}\right\vert_{#3}}

\def\expect#1#2{\underset{#1}{\mathbb{E}}\left[#2\right]}
\def\cov#1#2{\underset{#1}{\mathrm{Cov}}\left(#2\right)}

\def\abs#1{\left|#1\right|}

\DeclareMathOperator*{\argmin}{\mathrm{argmin}}

\def\p{\mathcal{P}}

\def\thetaopt{\hat{\theta}_{\mathrm{opt}}}
\def\thetasamp{\hat{\theta}_{\mathrm{samp}}}
\def\thetahat#1{\hat{\theta}_{\mathrm{#1}}}
\def\alphahat{\hat{\alpha}}

\def\X{X}

\def\w{w}
\def\v{v}

%% file: abstract.tex
One can typically form a local robustness metric for a particular problem quite
directly, in Markov chain Monte Carlo (MCMC) applications as well as VB.
However, we argue that simply forming a local robustness metric is not enough:
the hard work is showing that it is useful. Computability, interpretability, and
the ability of a local robustness metric to extrapolate well, are more important ---
and more difficult to establish --- than mere computation of derivatives.

%% file: introduction.tex
We feel very grateful to have our work carefully read and commented on by so
many insightful respondents. We would like to thank Professor Steel and the
editorial board of Bayesian Analysis for selecting our work and making this
discussion possible.  Statistical robustness is a venerable topic of
conversation and we have no doubt that the present discussion will continue far
into the future.

We might roughly categorize points made in the responses as
follows:\footnote{Please think of these categories as atom locations from an
Indian buffet process, not a Chinese restaurant process; respondents engaged at
times with multiple categories simultaneously.  For example, \gilleyva{} ask how
to form a VB approximation to an exchangeable stick breaking (ESB) prior (item
2) in order to form a local sensitivity metric (item 3) to assess whether the
ESB prior is robust for pointwise density estimation (item 1).}
\begin{enumerate}
\item Would a different model or summary statistic be more or less
robust than the Dirichlet process (DP) and number of clusters?
\item Can (or should) one form variational Bayes (VB) approximations to
different models from the BNP literature?
\item Can one form a local sensitivity metric to different quantities of
interest, modeling or fitting parameters, different posterior
approximation procedures, or some combination of all of these?
\end{enumerate}

Questions in category (1) and (2) are natural and important, since our work is
based on arguably the most canonical Bayesian nonparametric prior (the DP), and
a fairly vanilla VB approximation (a mean field approximation to a truncated
stick-breaking representation).  The evaluation of our robustness ideas with
respect to a wider range of priors and approximations is certainly warranted.
Nevertheless, in the present rejoinder we will focus on questions in category
(3), largely because we feel that our use of local sensitivity metrics
constitutes our work's most distinctive contribution.

Can one form a local robustness metric for a particular problem?  In
\cref{sec:derivatives} of the present rejoinder, we will argue: typically, yes,
quite directly, in Markov chain Monte Carlo (MCMC) applications as well as VB.
In \cref{sec:requested} we derive local robustness metrics for a select few
settings that were described in the responses.  After reading
\cref{sec:derivatives}, we hope that all readers of this rejoinder feel able and
empowered to form and investigate local robustness metrics for their own
particular problems.

However, in the subsequent \cref{sec:useful}, we will argue that simply forming
a local robustness metric is not enough: the hard work is showing that it is
useful. Computability, interpretability, and the ability of a local robustness
metric to {\em extrapolate} well, are more important --- and more difficult to
establish --- than mere computation of derivatives.  It is this work of
establishing usefulness that we have endeavored to undertake in the present
paper, and to which we wish to call attention as a foundation for further work.

As might be expected in a topic as established as robustness, the points made in
this rejoinder are not new, and have in fact been argued in the past by many of
our own respondents.  Nevertheless, we hope that by emphasizing the relative
ease of computing derivatives and the relative difficulty of showing their
utility in particular contexts, we can help advance the research agenda in this
important and challenging area.

%% file: derivatives.tex
A great deal of statistical inference---including Bayesian statistics---fixes some hyperparameter $\omega$ and then performs posterior inference using some combination of two types of estimators:
\begin{itemize}
\item The solution to a system of estimating equations:
$\thetaopt :={} \theta \textrm{ such that } G(\theta, \omega) = 0$.
\item A posterior moment from a density known up to a constant:
$\thetasamp :={} \expect{\p(\zeta \vert \omega)}{H(\zeta)}$.
\end{itemize}
%
%
An example of $\thetaopt$ could be the parameter that sets the gradient of a VB
loss function to zero (as in our paper), and an example of $\thetasamp$ could be
a posterior mean.  In practice, we may not be able to compute either exactly:
$\thetaopt$ might be approximated using numerical optimization, and $\thetasamp$
may be approximated using Markov chain Monte Carlo (MCMC).  Below, we will
briefly discuss the consequences of using approximations, but assume for the
moment that we can compute $\thetaopt$ and $\thetasamp$ to a desired accuracy.
A practitioner might then ask, ``what would happen if $\omega$ had taken on some
different value?'' The techniques of local robustness approximately answer this
question by forming a series approximation using derivatives of the maps $\omega
\mapsto \thetaopt(\omega)$ and $\omega \mapsto \thetasamp(\omega)$.

There are simple, general formulas for both these derivatives, under certain
common (but not universal) regularity conditions. For notational simplicity,
take $\omega$ to be a scalar for the moment. Furthermore, let us take
$\thetaopt(\omega)$ to be finite-dimensional, $\p(\zeta \vert \omega)$ to be
defined as a Radon-Nikodym derivative with respect to a common dominating
measure for all $\omega$, assume that we can exchange integration and
differentiation as needed, and assume all needed partial derivatives exist.
Then
\begin{align}
\fracat{d \thetaopt(\omega)}{d \omega}{\omega_0} ={}&
    \left(\fracat{\partial G(\theta, \omega)}
                 {\partial \theta}{\theta = \thetaopt(\omega_0),
                                   \omega=\omega_0} \right)^{-1}
    \fracat{\partial G(\theta, \omega)}
                {\partial \omega}{\theta = \thetaopt(\omega_0),
                                  \omega=\omega_0}
\label{eq:opt_deriv}
\end{align}
and
\begin{align}
\fracat{d \thetasamp(\omega)}{ d \omega}{\omega_0} ={}&
\cov{\p(\zeta \vert \omega_0)}{
    H(\zeta),
    \fracat{\partial \log \p(\zeta \vert \omega)}
           {\partial \omega}{\omega=\omega_0}}.
\label{eq:samp_deriv}
\end{align}
These two formulas have been noted many times in the literature, though we feel
it is worth calling attention to the simplicity of their form. The estimating
equation derivative in \cref{eq:opt_deriv} is formed using the implicit function
theorem \citep{krantz:2012:implicit} and is used, explicitly or implicitly, in
many local robustness works \citep{hampel:1974:influence,thomas:1989:assessing,hattori:2009:ipcqcasedeletion,shi:2016:gmmcasedeletion}, as well as our own present paper.  The
sampling derivative in \cref{eq:samp_deriv} is formed by differentiating under
the integral using a dominated convergence theorem
\citep[][Theorem 5.4]{billingsley:2008:probability} and appears widely, in various forms, in
the local Bayesian robustness literature and beyond
\citep{diaconis:1986:bayesconsistency,ruggeri:1993:infinitesimalposteriorsensitivity,gustafson:1996:localposterior,mohamed:2020:mcgradients}.

Importantly, \cref{eq:opt_deriv,eq:samp_deriv} can be
computed nearly automatically using automatic differentiation, using only the original solution $\thetaopt(\omega_0)$ or the ability
to compute moments of $\p(\zeta \vert \omega_0)$.
Furthermore, higher-order derivatives can be computed mechanically by repeatedly
applying \cref{eq:opt_deriv,eq:samp_deriv} to themselves.  For example,
higher-order versions of \cref{eq:opt_deriv} can be found in
\citet{giordano:2019:hoij}.
In practice, one can use a
corresponding numerical approximation to
$\thetaopt(\omega_0)$ or draws from $\p(\zeta \vert \omega_0)$ to approximate
the derivatives.
As observed by \griffin{}, the key practical difficulty with \cref{eq:opt_deriv}
is the solution of a linear system, and the key practical difficulty with
\cref{eq:samp_deriv} is Monte Carlo error.

One might contrast \cref{eq:opt_deriv,eq:samp_deriv}
with approaches that differentiate the optimization procedure or the sampling
procedure directly, as in chapter 6 of \citet{maclaurin:2016:autogradthesis} and
\citet{jacobi:2018:automatedMCMCsensitivity}, respectively,
both of which require
considerable bespoke computational effort, even with automatic differentiation.
The simplicity of \cref{eq:opt_deriv,eq:samp_deriv} comes at a cost, however, of
assuming (respectively) that $\thetaopt$ actually solves the estimating
equation, or that we are actually able to approximate draws from $\p(\zeta
\vert \omega_0)$. Studying \cref{eq:opt_deriv,eq:samp_deriv} in the presence of
violations of these assumptions is exciting and ongoing work, most notably in the setting of optimization~\citep{bae:2022:influencefunctionsanswerwhatquestion}.

Although \cref{eq:opt_deriv,eq:samp_deriv} apply to scalar $\omega$, they extend readily to multivariate and even
functional derivatives, since one can use scalar derivatives to differentiate
along a path in a multivariate space.  Different notions of ``derivative''
differ only in how they conceptually bundle these path derivatives together---as
a basis for a gradient in finite dimensional vector spaces
\citep{fleming:2012:functionsofseveralvariables}, as a basis for a tangent
plane in a geometric perspective \citep{mcinerney:2013:differentialgeometry,murray:1993:differentialgeometryandstatistics}, as Hadamard or
Fr{\'e}chet derivatives in infinite-dimensional spaces according to the
smoothness of the underlying function \citep{averbukh:1967:theory,zeidler:2013:nonlinearv1}.  In each case, however, for a particular path, the
derivative is always formally the same, and computable by
\cref{eq:opt_deriv,eq:samp_deriv}.

Given estimators of the form $\thetaopt$, $\thetasamp$, or some smooth
combination of them, the capacity to imagine a parameterized set of
perturbations of interest (and a class of univariate paths through it), one can
form local robustness measures---even for infinite-dimensional perturbations---using little more than \cref{eq:opt_deriv,eq:samp_deriv}, univariate calculus,
and the chain rule.  In the following section, we will demonstrate this point in a few
of the settings described by the respondents.

%% file: requests.tex
We now demonstrate our claim that derivatives are typically straightforward to
compute by doing so for several of the settings requested by our respondents:
case influence measures \citep{cook:1977:detectionofinfluential},
exchangeable stick breaking (ESB) processes \citep{gil:2021:stickbreakingesb},
empirical Bayes (EB) procedures \citep{mcauliffe:2006:nonparametriceb}, and
robustness to the loss function.  For ESB processes and EB we will discuss
some interesting challenges that arise from working with infinite-dimensional priors
in BNP models.

Many respondents noted that our (classical) approach to deriving local
robustness measures extends readily to other settings.
Readers who are similarly convinced that it is not difficult to
compute local robustness derivatives for a wide range of applications, for both
MCMC and optimization-based statistical procedures, can safely skip to
\cref{sec:useful}.

For this short rejoinder we have selected only a few settings from the
responses to address in detail, and we have chosen to prioritize the
settings that are most unlike the results in our paper.  Unfortunately,
doing so means forgoing discussion of some ideas which seem
particularly promising to us,
such as the proposal of \miscouridou{} to apply local robustness techniques
to the VB approximations for generalized gamma processes
given in \citet{lee:2016:finitevbbfry}.

Throughout this section, we will take $\zeta$ to denote all parameters of
a model and $X$ to denote observed data, so that the posterior is $\p(\zeta
\vert X)$. Let $\phi(\zeta)$ be some quantity of interest.
Note that, in \cref{eq:samp_deriv}, we need only differentiate
$\log \p(\zeta, X \vert \omega)$ rather than $\log \p(\zeta \vert X, \omega)$,
because the normalizing
constant $\p(\X \vert \omega)$ does not depend on $\zeta$ and does not
contribute to the covariance.

\paragraph{Case influence.}

\maceachern{} connect our work to a long history of frequentist and
Bayesian ``case influence'' literature.  This literature attempts to quantify the
importance of individual datapoints or groups of datapoints on a particular
inferential procedure.  \maceachern{} point to a set of works, beginning
 with \citet{cook:1977:detectionofinfluential}, which is particularly
concerned with ``outliers'' or ``gross errors'' as popularized by
\citet{huber:1964:robustlocation}.\footnote{A short historical account of this branch of
robust statistics is given by \citet{stigler:2010:robustnesshistory}.}
Indeed, the idea of using local approximations to robustness
under generic data perturbations goes back even farther---at least
as far as \citet{mises:1947:asymptotic}---and has been employed for asymptotic
theory~\citep{serfling:1980:approximation,
shao:2012:jackknife, vaart:1996:empiricalprocesses}, design and
analysis of robust estimators \citep{hampel:1974:influence}, approximation
of cross-validation in machine learning~\citep{koh:2017:blackboxinfluence,giordano:2019:swiss} and more.

To connect this broad literature to our work, we can augment each
datapoint with a scalar-valued weight, $\w_n$, in such a way that $\w_n = 1$
represents no change, and  $\w_n = 0$ represents omitting the datapoint from the
model. Specifically, letting $\w = (\w_1, \ldots, \w_N)$ and $\X = (\X_1,
\ldots, \X_N)$, we can write the log likelihood in a Bayesian model as
\begin{align*}
\log \p(\X, \zeta \vert \w) = \sum_{n=1}^{N} \w_n \log \p(\X_n \vert \zeta) + \log \p(\zeta),
\end{align*}
with $\p(\zeta \vert \X, \w)$ representing the corresponding posterior.  With
unit weights, we recover the original posterior:  $\p(\zeta \vert \X, \w=(1,
\ldots, 1)) = \p(\zeta \vert \X)$.  When $\w_n = 0$ but all other weights are
one, data point $n$ is left out.  Similarly, one can drop or replicate
any set of data points using the appropriate configuration of zeros, ones, or
other integers.

The advantage of writing $\log \p(\zeta \vert \X, \w)$  in this way is that we
can take a particular $\w_n$ to be our hyperparameter $\omega$ and apply
\cref{eq:opt_deriv,eq:samp_deriv} to form a local approximation to leaving out
(or replicating) sets of datapoints.  The form of the derivative for estimating equations
resulting from \cref{eq:opt_deriv} is the well-known empirical influence
function for $M$-estimators (see, e.g., eq.\ 2.3.5 of \citet{hampel:1986:robust}).
Perhaps less widely known is the corresponding result for MCMC estimators, which
is
\begin{align}
\fracat{\partial \expect{\p(\zeta \vert\X, \w)}{\phi(\zeta)}}
       {\partial \w_n}{\w=(1, \ldots, 1)} ={}&
       \cov{\p(\zeta \vert \X)}{\phi(\zeta), \log \p(\X_n \vert \zeta)}.
\label{eq:bayes_infl}
\end{align}
The right-hand side of \cref{eq:bayes_infl} can be estimated from MCMC
samples.  Then the quantity given in  \cref{eq:bayes_infl}
is precisely the ``Bayesian empirical influence function,'' evaluated at $\X_n$,
for the statistic $\expect{\p(\zeta \vert \X)}{\phi(\zeta)}$.
As with the frequentist influence function, \cref{eq:bayes_infl} may be used
to approximate all sorts of case deletion schemes from both the frequentist and
Bayesian literature---as long as one can show that it provides a good
approximation to the effect of actually removing the points.\footnote{
We should note that a first-order approximation is inadequate when taking some
form of KL divergence from the original posterior as the quantity of interest,
as is done in much of the literature cited by \maceachern{} \citep[e.g.,][]{johnson:1983:predictiveinfluence,mcculloch:1989:localbayesianmodelinfluence,carlin:1991:expectedutilityinfluence,thomas:2018:reconcilingcurvatureandis}.  This KL divergence is minimized
at $\w = (1, \ldots, 1)$, so the first derivative with respect to the weights is
zero, and one must form a local approximation using a second-order derivative.
However, all our comments in the present rejoinder, particularly
\cref{sec:useful}, apply to local second-order approximations as well as to
first-order approximations.
}

\paragraph{Dirichlet-driven ESB models.}

\gilleyva{} ask about local sensitivity analysis for Dirichlet-driven
exchangeable stick breaking (ESB) models \citep{gil:2021:stickbreakingesb}.  The
joint stick distribution in an ESB model is controlled by a parameter $\rho \in
[0,1]$ that smoothly transitions between stick-breaking priors with independent
and identically distributed sticks and stick-breaking priors for which all the
sticks take a common value.  \gilleyva{} take the quantity of interest to be the
posterior estimate of the density of the data generating distribution
evaluated at a point---a quantity which we can call $\phi(\zeta)$---and ask how
$\expect{\p(\zeta \vert X, \rho)}{\phi(\zeta)}$ depends on $\rho$.  \gilleyva{}
run an MCMC chain, but then, in order to compute local robustness measures,
attempt to construct a VB approximation to this posterior, observing that one
would need either to make a (potentially limiting) mean field assumption on the
sticks or deal with a computationally intractable normalizing constant.

We will avoid the question of how to construct a VB approximation in their
setting, and derive instead a local sensitivity measure that can be used with an
MCMC chain---as long as the stick-breaking distribution can be effectively
truncated at $K$ sticks for some finite $K$. Let the truncated stick lengths be
denoted by $\v_1, \ldots, \v_K$.  We can imagine several ways to truncate an ESB
model, but for the present purposes, one would need to be able to sample from
the truncated model, and the prior $\p(\v_1, \ldots \v_K \vert \rho)$ would need
to be tractable and smooth for any draw from the MCMC chain.\footnote{Note that
for the un-truncated Dirichlet-driven ESB model, the density of any finite
number of sticks is tractable and smooth as a function of $\rho$: see Appendix
E, Section 5 of the supplementary material to \citet{gil:2021:stickbreakingesb}
where the needed density is derived as part of a Gibbs sampler for $\rho$.}
By \cref{eq:samp_deriv} we then have
\begin{align}
\fracat{\partial \expect{\p(\zeta \vert \X, \rho)}{\phi(\zeta)}}
       {\partial \rho}{ \rho = \rho_0}
={}&
\cov{\p(\zeta \vert \X, \rho_0)}
    {\phi(\zeta),
    \fracat{\partial \log \p(\v_1, \ldots, \v_K \vert \rho)}
           {\partial \rho}{\rho=\rho_0}}.
\label{eq:esb_deriv}
\end{align}
The preceding sampling covariance can in principle be
approximated from MCMC samples, with no need to form a VB approximation.

Carefully
considering the implications of truncating ESB models is beyond the scope
of this rejoinder, but it is worth noting the challenges for local
robustness if one does not truncate, especially since
\citet{gil:2021:stickbreakingesb} use a slice sampler
and do not truncate the stick-breaking distribution.
If the log prior contained an infinite
number of terms, it is not obvious that one could apply the
dominated convergence to derive \cref{eq:samp_deriv}.
There exist sampling
schemes that in fact sample only a finite number of sticks without truncation;
see, e.g., \citet{gil:2021:stickbreakingesb} and \citet{walker:2007:sampling}. Similarly,
one might hope that one could apply \cref{eq:samp_deriv} without truncation
by conditioning on auxiliary random variables in the
$\log \p(\zeta \vert \omega)$ term in \cref{eq:samp_deriv}.
But this term cannot be conditional on quantities that are
random in $\p(\zeta \vert \omega)$.
Additionally, the truncation will,
in general, have an effect; here, the quantity of interest
(the posterior estimate of the data density at a point) has
nonzero correlation with \emph{all} sticks.
Although the posterior density at a point plausibly has diminishing
correlation with sticks
that come later in the process, one could design adversarial quantities of
interest---e.g., the value of the 10,000-th stick---for which
truncation would be quite inaccurate.  Developing tractable sensitivity
measures for infinite-dimensional posteriors is an interesting problem,
though we suspect that \cref{eq:esb_deriv} will still be informative using straightforward finite
truncation, especially given that any error
in the linear approximation may well be larger than the error induced
by truncation.

\paragraph{Empirical Bayes.}

\rebaudo{} ask whether Empirical Bayes (EB) methods for setting a DP prior
might be more robust. One might answer such a question
empirically by forming local robustness measures for EB procedures, which we
will now undertake.

Concretely, \citet{mcauliffe:2006:nonparametriceb} rely on an EB procedure to
choose the DP concentration parameter.  Their EB procedure takes the following
general form. Fix some hyperparameter $\omega$, which might be a case weight
(see above), a perturbation of the base measure, etc.
The EB procedure then finds a prior
parameter $\alphahat$ that satisfies, for some $F$ and $G$,
\begin{align}
G(\alphahat, m(\alphahat, \omega)) ={}& 0 \quad \textrm{ where }\quad
    m(\alpha, \omega) :={} \expect{\p(\zeta \vert \X, \alpha, \omega)}{F(\zeta)}.
\label{eq:eb}
\end{align}
Specifically, to set the concentration parameter of a DP prior using EB,
\citet{mcauliffe:2006:nonparametriceb} take $\alpha$ to be the DP concentration
parameter, $F(\zeta)$ to denote the number of clusters observed for the dataset
$\X$ of size $N$, and $G(\alpha, m) = \sum_{n=1}^N \frac{\alpha}{\alpha + n - 1} -
m$ (see their eq. 8).

EB procedures such as this one take the form of an estimating equation that
depends on a posterior moment, which can be differentiated by
\cref{eq:opt_deriv,eq:samp_deriv} and the chain rule.  Note that $\alphahat$
depends on $\omega$, which we write as $\alphahat(\omega)$. Additionally, write
$\alphahat_0 := \alphahat(\omega_0)$ and $m_0 := m(\alphahat_0, \omega_0)$.
To compute a local robustness measure for the posterior expectation of
$\phi(\zeta)$, we must compute
\begin{align*}
\fracat{\partial \expect{\p(\zeta \vert \X,
\alphahat(\omega), \omega)} {\phi(\zeta)}} {\partial \omega}{\omega_0},
\end{align*}
accounting for the $\omega$ dependence in both the empirical Bayes procedure and
the final posterior expectation.
This derivative can be readily computed by applying the chain rule and
\cref{eq:opt_deriv,eq:samp_deriv}. The result, given below in
\cref{eq:eb_deriv}, is a bit tedious, but its computation is entirely mechanical
(and automatable) and applies to any EB procedure of the form in
\cref{eq:eb}.\footnote{For compactness, we have suppressed some evaluation
notation in this display; for example, we write $\alphahat_0$ in place of
$\alpha = \alphahat_0$.  The evaluation is always done for the parameter with
respect to which we are differentiating.}
\begin{align}
\fracat{\partial \expect{\p(\zeta \vert \X, \alphahat(\omega), \omega)}
                        {\phi(\zeta)}}
       {\partial \omega}{\omega_0}
       ={}&
\cov{\p(\zeta \vert \X, \alphahat_0, \omega_0)}
    {\phi(\zeta),
    \fracat{\partial \log \p(\zeta, \X \vert \alphahat_0, \omega)}
           {\partial \omega}{\omega_0} +
    \fracat{\partial \log \p(\zeta, \X \vert \alpha, \omega_0)}
           {\partial \alpha}{\alphahat_0}
        \fracat{d \alphahat(\omega)}{d \omega}{\omega_0}
    }
\nonumber \\
\textrm{where }
\fracat{d \alphahat(\omega)}{d \omega}{\omega_0} ={}&
-\left(
    \fracat{\partial G(\alpha, m(\alpha, \omega_0))}
           {\partial \alpha}{\alphahat_0}
\right)^{-1}
\left(
     \fracat{\partial G(\alphahat_0, m)}{\partial m}{m_0}
     \fracat{\partial m(\alphahat_0, \omega)}{\partial \omega}{\omega_0}
\right),
\nonumber \\
\fracat{\partial G(\alpha, m(\alpha, \omega_0))}{\partial \alpha}{\alphahat_0}
={}&
    \fracat{\partial G(\alpha, m_0)}{\partial \alpha}{\alphahat_0} +
    \fracat{\partial G(\alphahat_0, m)}{\partial m}{m_0}
    \fracat{\partial m(\alpha, \omega_0)}{\partial \alpha}{\alphahat_0},
\nonumber \\
\fracat{\partial m(\alpha, \omega_0)}{\partial \alpha}{\alphahat_0} ={}&
\cov{\p(\zeta \vert \X, \alphahat_0, \omega_0)}
    {F(\zeta),
    \fracat{\partial \log \p(\zeta, \X  \vert \alpha, \omega_0)}
                         {\partial \alpha}{\alphahat_0}},
\nonumber \\\textrm{and }
\fracat{\partial m(\alpha_0, \omega)}{\partial \omega}{\omega_0} ={}&
\cov{\p(\zeta \vert \X, \alphahat_0, \omega_0)}
    {F(\zeta),
    \fracat{\partial \log \p(\zeta, \X  \vert \alphahat_0, \omega)}
                         {\partial \omega}{\omega_0}}.
\label{eq:eb_deriv}
\end{align}

One might ask whether \cref{eq:eb_deriv} can be applied to the EB procedure used
by \citet{mcauliffe:2006:nonparametriceb} for the base measure.  Unfortunately,
since \citet{mcauliffe:2006:nonparametriceb} estimate the base measure
nonparametrically, the space of possible base measures is infinite-dimensional,
and so one cannot apply \cref{eq:opt_deriv} directly. Versions of
\cref{eq:opt_deriv} for infinite-dimensional parameters exist~\citep[see, e.g.,
Chapter 4 of][]{zeidler:2013:nonlinearv1}, though applying them in practice
seems to be challenging and beyond the scope of this short
rejoinder.  Alternatively, one could represent the base measure using a
large but finite basis and apply \cref{eq:eb_deriv}.

\paragraph{Loss functions.}

\maceachern{} ask whether we can compute sensitivity to the loss
function in a Bayesian analysis.  Formally, for a posterior
$\p(\zeta \vert X)$ and loss function $L$,
under common regularity conditions we have
\begin{align}
\thetahat{loss} :=
    \argmin_{\theta} \expect{\p(\zeta \vert \X)}{L(\zeta, \theta)}
\quad\Leftrightarrow
\expect{\p(\zeta \vert X)}{
\fracat{\partial L(\zeta, \theta)}
       {\partial \theta}{\theta=\thetahat{loss}}} = 0.
\label{eq:bayes_loss}
\end{align}
We will consider the common situation described in \cref{eq:bayes_loss}.
However, we note that, by exchanging the order of local robustness derivatives
and posterior expectations, this approach could be naturally extended to
estimators of the form given in \citet{lee:2014:bayesianinferencefunctions}, i.e.,
$\expect{\p(\zeta \vert \X)}{\argmin_{\theta} \int L(y,
\theta) \zeta(d y)}$ for a distribution-valued $\zeta$.

\Cref{eq:bayes_loss} defines an estimating equation for $\thetahat{loss}$. We
can parameterize a path to a different loss function using $L(\zeta, \theta,
\omega) = L(\zeta, \theta) + \omega \Delta(\zeta, \theta)$ for some
$\Delta(\zeta, \theta)$. Then,  apply \cref{eq:opt_deriv} to
\cref{eq:bayes_loss}, and interchange differentiation and integration to get
\begin{align}
\fracat{d \thetahat{loss}(\omega)}{d \omega}{\omega = 0} ={}&
-\left(
\expect{\p(\zeta \vert X)}{
\fracat{\partial^2 L(\zeta, \theta)}{\partial \theta \partial \theta}
    {\theta=\thetahat{loss}}}
\right)^{-1}
\expect{\p(\zeta \vert X)}{
\fracat{\partial \Delta(\zeta, \theta)}{\partial \theta}
    {\theta=\thetahat{loss}}}.
\label{eq:loss_deriv}
\end{align}
For example, to estimate the effect of replacing the mean with the median,
we could take $L(\zeta, \theta)=\frac{1}{2}(\zeta - \theta)^2$,
$\Delta = \abs{\zeta - \theta} - L(\zeta, \theta)$, and
\begin{align}
\MoveEqLeft
\textrm{Median}(\p(\zeta \vert \X)) - \expect{\p(\zeta \vert \X)}{\zeta} ={}
\thetahat{loss}(1) - \thetahat{loss}(0)
\approx \fracat{d \thetahat{loss}(\omega)}{d \omega}{\omega = 0} (1 - 0)
\nonumber\\={}&
\expect{\p(\zeta \vert X)}
    {\ind{\zeta > \expect{\p(\zeta \vert \X)}{\zeta}}} -
\expect{\p(\zeta \vert X)}
    {\ind{\zeta < \expect{\p(\zeta \vert \X)}{\zeta}}}.
\label{eq:mean_med_approx}
\end{align}
For example, this approximation reasonably asserts that the median will exceed
the mean when the posterior is asymmetric, with a greater mass to the right of
the mean than to the left.  (But we will discuss some of its limitations in
\cref{sec:useful} below.)

Since $\Delta$ could be any function satisfying basic regularity conditions, one
could in principle use \cref{eq:loss_deriv} to explore the space of loss
functions---if one can believe that the approximation provided by $\Delta
\mapsto \fracat{d \thetahat{loss}(\omega)}{d \omega}{\omega = 0}$ is a good one
uniformly over the candidate set of perturbations $\Delta$.  However, again, we
do not necessarily recommend this for this particular path through the space of
loss functions.  On the contrary, we will use this example in \cref{sec:useful}
below as an example of a derivative that may not serve its intended purpose very
well.

%% file: useful.tex
In \cref{sec:requested} above, we derived local robustness measures for several
settings requested by our respondents, and we expect that readers can readily
derive most of the rest for themselves.  Have we solved all their problems?
Certainly not!  To the contrary, we will argue that the computation of
derivatives is straightforward, but showing their utility is harder.

In our view, a useful derivative should (at least) satisfy a few related
``usefulness desiderata'': 
(1) be readily computable to the desired accuracy, (2) be easily
interpretable, and, most importantly, (3) extrapolate well so as to
provide a reasonable approximation to the ``global robustness'' problem.  We have endeavored to show that certain derivatives are at least plausibly
useful, according to these criteria, for DP priors in VB approximations.  In
addition to considering Fr{\'e}chet differentiability---which is, arguably, a
rather low bar for a derivative to pass---we primarily demonstrated our local
robustness measure's ability to extrapolate through careful experiments and
comparison with refitting.  In certain situations, such as case influence in
large datasets, one can sometimes prove good extrapolation by bounding the
second derivative under readily interpretable conditions
~\citep[as in][]{giordano:2019:swiss}.


Evaluating the usefulness desiderata for the derivatives given in
\cref{sec:requested} above is worth doing, and it is the work which
constitutes most of the effort.  We do not believe that all the
results of \cref{sec:requested} will pass the test.  Let us focus on the
loss function example, though many of these potential problems apply to the
other settings as well.

\textbf{Computability.}  The expected Hessian inside the inverse
in \cref{eq:loss_deriv} will have Monte Carlo error if estimated with MCMC,
which will bias the inverse.  Furthermore, if the difference between
$L$ and $L + \Delta$ occurs mostly in the tails, the expectation of the
derivative of $\Delta$ may suffer from high MCMC noise.

\textbf{Interpretability.}  The loss function derivative in \cref{eq:loss_deriv}
will behave pathologically as an approximation to losses that are pointwise
close to the original loss function, but have very large derivatives near
$\thetahat{loss}$. Without a judiciously constrained search space to exclude
such alternative loss functions, \cref{eq:loss_deriv} will provide poor
guidance when exploring the space of alternative loss functions. Unfortunately,
a more complex search space comes at a cost, which is the computational
difficulty of optimizing a linear form (i.e., the derivative in
\cref{eq:loss_deriv}, viewed as a functional of $\Delta$) over this space.

\textbf{Extrapolation.}  The example of the mean and median shows that the
derivative \cref{eq:loss_deriv} may not always extrapolate well in common use
cases. Though we know that the mean and median can be arbitrarily different in
general, the approximation to the difference in \cref{eq:mean_med_approx} cannot
be larger in magnitude than one.

Attempting to investigate and repair these deficiencies---e.g., by
improved MCMC sampling, alternative paths through the space of loss functions,
and the selection of search sets in the space of loss functions---is an interesting and valuable project, and one that
may require considerably more effort than derivation of the local
robustness approximation.


The usefulness desiderata sufficed for our objective, which was primarily to
provide a tool for quickly exploring the space of stick-breaking priors in a way
that is not too computationally or technically burdensome, for a particular
quantity of interest.  On the other hand, we do not attempt to detect
sensitivity of the entire model (e.g., with a whole-distribution
divergence measure), we
do not assert that a large worst-case derivative implied non-robustness 
(e.g., the corresponding prior may have looked
subjectively unreasonable), and we do not assert that our good results on
extrapolation will necessarily hold in very different settings (we primarily
showed good extrapolation via experiment).  In this sense, our objectives are
somewhat different than much of the classical robustness literature, which we
see as attempting to provide more universal notions of ``robustness.''   For
example, the foundational works of
\citet{ruggeri:1993:infinitesimalposteriorsensitivity}, \citet{basu:1996:local},
\citet{gustafson:1996:localposterior}, to which we are quite indebted, appear to
take as their task the \emph{definition} of a single number which can be
interrogated, relatively free of context, to ascertain whether a posterior is
``robust'' or not.  This goal is reflected in how their techniques are used, for
example, in \citet{basu:2000:robustnessbnp}.  A similar goal of finding universal
metrics of ``data importance'' motivates much of the case deletion literature;
it is perhaps for this reason that many authors focus on various forms
of whole-model KL divergence or likelihood ratios~\citep[see, e.g.,][]{johnson:1983:predictiveinfluence,cook:1986:assessment,carlin:1991:expectedutilityinfluence}.  The production of a universally
valid local robustness metric requires even stricter conditions on the
derivative than our usefulness desiderata; e.g., it must extrapolate well
in all directions, its worst-case perturbation must lead to a subjectively reasonable
prior, the researcher must actually care about whole-model sensitivity
and not just a particular posterior quantity, and so on.

Our context-specific approach and a more universalist approach need not be at
odds.  On the contrary, the intuition and best practices arising from routine
and systematic assessment of prior assumptions might lead, in the end, to better
and more universally applicable metrics of robustness.  Similarly, asymptotic
analysis of the sort advocated by \ascolani{} can inform and be informed by the
robustness of particular finite-data settings.

Where local approximations fail to satisfy the usefulness desiderata, there is
ample room for creativity.  The analysis of \griffin{}, which both clearly
demonstrates a failure of a linear approximation to extrapolate and suggests a
solution, seems exemplary to us. Their idea of performing sensitivity analysis
separately for a number of local modes seems promising; to their suggestion we
might add forming a (second order) approximation to the value of the ELBO at
these modes as well, in order to assess when the relative ordering of the modes
changes.  Local approximations might also augment other schemes, such as
suggesting quadratic transforms for the importance sampling techniques of
\citet{maceachern:2000:importancelinkMCMC}.

We hope that researchers will feel empowered by this work to creatively explore
the space of model perturbations, relatively unencumbered by the difficulty of
deriving local robustness measures, but attentive to their ability to answer
useful questions in their own modeling contexts.